\documentclass[a4paper,10pt,twocolumn,preprint,3p]{elsarticle}
\usepackage{geometry}
\usepackage{amsmath}
\usepackage{amssymb}
\usepackage[usenames,dvipsnames]{color}
\usepackage{hyperref}

\newcommand{\okk}{\mathcal O(M^{-2})}
\newcommand{\id}{1}
\newcommand{\Mink}{\mathcal M_{1,3}}

\hypersetup{
    colorlinks=true,         
    linkcolor=red,          
    citecolor=red,        
    urlcolor=Violet             
}

\begin{document}

  \begin{frontmatter}
\author{\textbf{Stjepan Meljanac}$^a$, \textbf{Daniel Meljanac}$^a$, \textbf{Flavio Mercati}$^{b}$ and \textbf{Danijel Pikuti\'c}$^a$
\vspace{12pt}
\\
$^a$Division of Theoretical Physics, Rudjer Bo\v{s}kovi\'c Institute, Bijeni\v{c}ka c. 54, HR-10002 Zagreb, Croatia
\\
$^b$Perimeter Institute for Theoretical Physics, 31 Caroline St. N.,
\\
Waterloo, ON, N2L~2Y5  Canada}
\title{\sc{Noncommutative Spaces and Poincar\'e Symmetry}}
\begin{abstract}

We present a framework which unifies  a large class of non-\-com\-mu\-ta\-ti\-ve spacetimes that can be described in terms of a deformed Heisenberg algebra. The commutation relations between spacetime coordinates are up to linear order in the coordinates, with structure constants depending on the momenta plus terms depending only on the momenta. The possible implementations of the action of Lorentz transformations on these deformed phase spaces are considered, together with the consistency requirements they introduce. It is found that Lorentz transformations in general act nontrivially on tensor products of momenta. In particular the Lorentz group element which acts on the left and on the right of a composition of two momenta is different, and depends on the momenta involved in the process. We conclude with two representative examples, which illustrate the mentioned effect.
\end{abstract}
  \end{frontmatter}

\section{Introduction}

The framework of Quantum Field Theory\footnote{Or rather Effective Field Theory.} (QFT) has been extremely successful in predicting new phenomena and reaching great accuracy in the description of known phenomena. This framework, however, is based on a fixed kinematic structure given by the background spacetime on which the matter fields propagate. In the most successful applications of QFT this background is assumed to be Minkowski space $\mathcal M_{1,3}$. Such a structure is untenable if we want our description of physics to be valid up to energies of the order of the Planck scale $E_p \sim 10^{28}eV$ . At these energies gravitational effects become important, and a naive effective description involving quantum fluctuations of the gravitational field around a Minkowski background turns out not to be renormalizable~\cite{FeynmanLecturesGravitation}, signalling that the theory needs an unknown ultraviolet completion.

A possible insight towards progress comes from 2+1 dimensional Quantum Gravity, which is a topological theory without local, propagating degrees of freedom. If this theory is coupled to a scalar field, and the topological degrees of freedom of the metric are integrated away, one ends up with an effective theory for the scalar field on a background which is not Minkowski space~\cite{MatschullWelling,FreidelLivine}. The background geometry is noncommutative, in the sense that the algebra of functions over this background is not abelian.\footnote{For the meaning of noncommutative spacetimes, read~\cite{MajiIntroToNCSpaces}.}

This opens up the possibility that the sought-after UV completion of Quantum Gravity might not be found by simply changing the field content or the symmetries within the traditional scheme of QFT on a classical background. It may be necessary to generalize QFT to \emph{Noncommutative} QFT (NCQFT), in which the kinematical structure is not simply given by a spacetime background, but it is encoded in some nontrivial commutation relations.

It becomes then necessary to understand the types of noncommutative spaces that are the candidates to play the role of background for NCQFT, the kinematic structures they encode, and their symmetries. Usually a noncommutative spacetime is defined by specifying the commutation relations for a coordinate basis, $\hat x_\mu$, $\mu = 0,...,3$. The three most popular types of commutation relations considered in the literature are (i) Lie-algebra type~\cite{kPoincarePLB92}, (ii) Moyal type~\cite{Groenewold,Moyal} and (iii) Snyder type~\cite{Snyder}:
\begin{align}
&(i)&[\hat x_\mu, \hat x_\nu]&=i\hat x_\alpha C_{\mu\nu}{}^\alpha, && C_{\mu\nu}{}^\alpha \in \mathbb R && \nonumber\\
&(ii)&[\hat x_\mu, \hat x_\nu]&=i\Theta_{\mu\nu}, && \Theta_{\mu\nu} \in \mathbb R &&\label{nc-types} \\
&(iii)&[\hat x_\mu, \hat x_\nu]&=isM_{\mu\nu}, && s \in \mathbb R && \nonumber
\end{align}
where $M_{\mu\nu}$ is the generator of Lorentz transformations.

In this paper we introduce a framework which includes all the cases of eq. \eqref{nc-types} as sub-cases, and is capable of describing much more general cases. This framework is based on the idea that any non-commutative algebra can be obtained as a particular nonlinear realization of the Heisenberg algebra.

\section{Deformed phase space}

We start with the Heisenberg algebra $\mathcal H$, which is a unital algebra generated by the eight generators $x_\mu$, $p_\nu$, $\mu,\nu =0,1,2,3$. The commutation relations are $[x_\mu, x_\nu]=0$, $[p_\mu, p_\nu]=0$,  $[p_\mu, x_\nu]=- i \eta_{\mu\nu}$, where  $\eta = \text{diag} (-1,1,1,1)$.

A nonlinear change of basis allows us to define a set of generically noncommutative coordinates:
\begin{equation}\label{NoncommutativeCoordinates}
\hat x_\mu =  \textstyle x_\alpha \varphi^\alpha{}_\mu \left( \frac p M \right) +\frac1M\chi_\mu \left( \frac p M \right)  ,
\end{equation}
where  $M$ is a constant with the dimensions of a mass which parametrizes the magnitude of spacetime noncommutativity (and is expected to be of the order of the Planck mass). Assuming that at large distances (or low energies) the effects of noncommutativity become irrelevant, and the generators (\ref{NoncommutativeCoordinates}) coincide with $x_\mu$, we can impose the `boundary conditions' $ \varphi^\alpha{}_\mu \left(0\right) = \delta^\alpha{}_\mu$ and $\chi_\mu \left( 0 \right) =0$. If we assume $\varphi^\alpha{}_\mu $ to be an invertible matrix, the inverse relations to (\ref{NoncommutativeCoordinates}) exist: $x_\mu = \left( \hat x_\alpha - \chi_\alpha \right) (\varphi^{-1})^\alpha{}_\mu$.

Changing base from $x_\mu$ to $\hat x_\mu$ leads to the deformed Heisenberg algebra $\hat{\mathcal H}$:
\begin{equation}\label{DeformedHeisenbergAlgebra}
\begin{aligned}\textstyle
[\hat x_\mu ,\hat x_\nu ] = \frac i M  \hat x_\alpha  C_{\mu \nu}{}^\alpha\left( \frac p M \right)  +  \frac i {M^2} \Theta_{\mu \nu}\left( \frac p M \right) ,
\\ \textstyle
[ p_\mu , \hat x_\nu ] = - i  \varphi_{\mu\nu} \left( \frac p M \right)    ,~~~
[ p_\mu , p_\nu ] = 0 ,
\end{aligned}
\end{equation}
where $C_{\mu\nu}{}^\alpha = (\varphi^{-1})^\alpha{}_\beta \left( \varphi^\gamma{}_\mu  \partial_\gamma \varphi^\beta{}_\nu  -  \varphi^\gamma{}_\nu  \partial_\gamma \varphi^\beta{}_\mu \right)$ and
 $\Theta_{\mu\nu} =-\chi_\alpha (\varphi^{-1})^\alpha{}_\beta \left( \varphi^\gamma{}_\mu  \partial_\gamma \varphi^\beta{}_\nu  -  \varphi^\gamma{}_\nu  \partial_\gamma \varphi^\beta{}_\mu \right) -  \varphi^\gamma{}_\mu  \partial_\gamma \chi_\nu + \varphi^\gamma{}_\nu  \partial_\gamma \chi_\mu$,  where $\partial_\gamma=M\frac\partial{\partial p^\gamma}$. 
The generalized Jacobi relations are satisfied by construction. The commutators \eqref{DeformedHeisenbergAlgebra} of $\hat x_\mu$ unifiy into a single formalism the three kinds of noncommutative spaces which have been considered in the literature, listed in eq. \eqref{nc-types}.

From now on we will work in units in which $M=1$.

We can define the subalgebra  $\mathcal A \in \mathcal H$ of commutative coordinates as $\mathcal A = \text{span}(x_\mu)$. We can define a left-action of $\triangleright :\mathcal H \otimes \mathcal A \to \mathcal A  $ with the following axioms~\cite{EPJC}:
\begin{equation}
\begin{aligned}
&f(x) \triangleright g(x) = f(x) g(x), 
\\
&p_\mu \triangleright f(x) = [p_\mu , f(x)] \triangleright 1, 
\\
&p_\mu \triangleright 1 = 0,
\end{aligned}
\end{equation}
for all $f(x),g(x) \in \mathcal A$. This left-action corresponds to an action by multiplication of coordinates:
 $x_\mu \triangleright f(x) = x_\mu f(x)$, and an action of momenta by left-derivative:  $p_\mu \triangleright f(x) = - i \frac{\partial f}{\partial x^\mu}$. The action $\triangleright$ is an left algebra homomorphism, which means that it respects the product of $\mathcal H$ so that $(a b) \triangleright f = a \triangleright  ( b \triangleright f)$. So any product of $p_\mu$'s and $x_\mu$'s in a particular order will act on $f$ by applying derivatives and left-multiplications in the corresponding order. 

It can be proven explicitly~\cite{EPJC,HermitianRealizations,SIGMA} that 
\begin{equation}\label{ExpActionOn1}
e^{i k \cdot \hat x} \triangleright 1 = e^{i K(k) \cdot x + i g(k)}, 
\end{equation}
where $k \in \Mink$ and  $K :  \Mink \to  \Mink$ is an invertible map $K_\mu(K^{-1}(k))=k_\mu$ and $g :  \Mink \to \mathbb{R}$. 

For any function $f(x) \in \mathcal A$ that can be Fourier-transformed $f = \int d^4 k \tilde f(k) e^{i k \cdot x}$ we 
can define, using relation~(\ref{ExpActionOn1}), an element $\hat f \in \mathcal H$ such that $\hat f \triangleright 1 = f(x)$. This allows us to introduce a noncommutative star-product  $\star : \mathcal A \otimes \mathcal A \to \mathcal A$:
\begin{equation}
f(x)  \star g(x) =  \left(\hat f  \hat g \right) \triangleright 1 = \hat f  \triangleright g (x).
\end{equation}
The $\star$-product beween exponentials is \cite{KovMelJPA}
\begin{equation}
e^{i k \cdot x} \star e^{i q \cdot x} = e^{i K^{-1}(k) \cdot \hat x} \triangleright e^{i q \cdot x}.
\end{equation}
This is used to determine the composition law of momenta, which is physically relevant as it determines how particle momenta are conserved in vertices.

\section{Deformed composition of momenta}

With an explicit calculation, one can prove the following relation (see~\cite{HermitianRealizations,SIGMA})
   \begin{equation}\label{ExpActionOnExp}
 e^{i k \cdot \hat x} \triangleright e^{i q \cdot x} = e^{i\mathcal P(k,q) \cdot x + iQ(k,q)},
 \end{equation}
where, from Eq.~(\ref{ExpActionOn1}), we see that $K_\mu (k) = \mathcal P_\mu(k,0)$ and $g(k) = Q(k,0)$.
 Using the homomorphism property of $\triangleright$ we can deduce the following relation:
\begin{equation}\label{RelationForP}
e^{- i \lambda k \cdot \hat x} p_\mu e^{ i \lambda k \cdot \hat x} \triangleright e^{i q \cdot x}
= \mathcal P_\mu( \lambda k,q)  e^{i q \cdot x}.
\end{equation}
where $\lambda \in \mathbb R$. Differentiating such a relation with respect to $\lambda$ leads to
\begin{equation}\label{DiffEqForP}
\frac{d \mathcal P_\mu (\lambda k,q)}{d\lambda} = \varphi_\mu{}^\alpha \left( \mathcal P(\lambda k, q)\right) k_\alpha,
\end{equation}
notice that $\mathcal P_\mu$ only depends on $\varphi_\mu{}^\alpha$. We can determine $Q$ by differentiating wrt $\lambda$ the following modification of relation~(\ref{RelationForP}):
\begin{equation}\label{RelationForH}
\begin{aligned}
e^{- i \lambda k \cdot \left( x_\alpha \varphi^\alpha{}_\mu(p) \right) } p_\mu e^{ i \lambda k \cdot \hat x} \triangleright e^{i q \cdot x}
\\
=\mathcal P_\mu( \lambda k,q)  e^{i q \cdot x + i Q(\lambda k,q)},
\end{aligned}
\end{equation}
and using~(\ref{DiffEqForP}), one gets
\begin{equation}\label{DiffEqForH}
\begin{aligned}
\frac{d Q (\lambda k,q)}{d\lambda} = k_\alpha \chi^\alpha \left( P(\lambda k ,q)\right).
\end{aligned}
\end{equation}
We can see how $Q$ does depend on $\chi_\mu$, and it is zero if $\chi_\mu =0$.

The differential equations~(\ref{DiffEqForP}) and~(\ref{DiffEqForH}) determine $P_\mu$ and $Q$, if supplemented  with the boundary conditions $\mathcal P_\mu (k,0) = K_\mu (k)$, $\mathcal P_\mu (0,q) = q_\mu$, $Q(k,0) =g(k)$, $Q(0,q) = 0$, $K_\mu(0)=0$, $g(0) = 0$.
Using Eq.~(\ref{ExpActionOn1}) and (\ref{ExpActionOnExp}), we conclude that the star-product between two plane waves is
\begin{equation}
\begin{aligned}
& e^{i k \cdot  x} \star e^{i q \cdot x}  =  e^{i K^{-1}(k) \cdot \hat x -  i g \left( K^{-1}(k) \right)}  \triangleright e^{i q \cdot x}
\\
&= e^{i \mathcal P \left( K^{-1}(k),q \right) \cdot x + i Q \left( K^{-1}(k),q \right)-  i Q \left( K^{-1}(k),0 \right) }.
\end{aligned}
\end{equation}
The generalized addition rule for plane wave momenta is then
\begin{equation}
\left( k \oplus q \right)_\mu = \mathcal D_\mu (k,q) = \mathcal P_\mu \left( K^{-1}(k),q \right), 
\end{equation}
where  $\mathcal D_\mu(k,0) = k_\mu$ and $\mathcal D_\mu (0,q) = q_\mu$.

\section{Coproduct, twist and star product}

Let us introduce the coproduct of momenta~\cite{EPJC}
\begin{equation}\label{MomentumCoproduct}
\Delta p_\mu = \mathcal D_\mu \left( p \otimes \id , \id \otimes p \right),
\end{equation}
and the twist element~\cite{TwistedStatistics, IJMPA1305}
\begin{equation}\label{twist1}
\begin{aligned}
\mathcal F^{-1} &= : e^{  i \left( \id \otimes x^\mu \right)  (\Delta-\Delta_0)p_\mu+iG \left( p \otimes \id , \id \otimes p \right)} :
\\
&= : e^{  i \left( \id \otimes x^\mu \right) (\Delta-\Delta_0)p_\mu} : e^{iG \left( p \otimes \id , \id \otimes p \right)},
\end{aligned}
\end{equation}
where $ \Delta_0 p_\mu = p_\mu \otimes \id + \id \otimes p_\mu$ is the undeformed coproduct, $G(k,q)=Q(K^{-1}(k),q)-Q(K^{-1}(k),0)$, and $:\_:$ denotes a normal ordering prescription in which the coordinates $x_\mu$ stand at the left of the momenta $p_\mu$. The twist $\mathcal F^{-1}$ is determined up to the right ideal $\mathcal I_0$ defined by \cite{IJMPA1305, SIGMA2014}
\begin{equation}
m\left(\mathcal I_0(\triangleright\otimes\triangleright)(f\otimes g)\right)=0 \quad \forall f, g\in\mathcal A
\end{equation}
where $m : \mathcal H  \times \mathcal H  \to \mathcal H $ is the multiplication map of the algebra $ \mathcal H$.

The star-product is then defined as
\begin{equation}
\left( f \star g \right) (x) = m \left[ \mathcal F^{-1} \left( \triangleright \otimes \triangleright \right) \left( f \otimes g \right) \right]. 
\end{equation}
Note that if $[\hat x_\mu,\hat x_\nu]=0$, equation \eqref{DeformedHeisenbergAlgebra}, then the star product is commutative and associative but generally non-local. If $C_{\mu\nu}{}^\alpha\left(\frac pM\right)$ or $\Theta_{\mu\nu}\left(\frac pM\right)$ depend on momenta, then the star product is non-associative.

The following identities hold:
\begin{equation}
\hat x_\mu = m  \left[ \mathcal F^{-1} \left( \triangleright \otimes \id \right) \left( x_\mu \otimes 1 \right) \right],
\end{equation}
and
\begin{equation}
\hat f = m  \left[ \mathcal F^{-1} \left( \triangleright \otimes \id \right) \left( f  \otimes 1 \right) \right], ~ \hat f \triangleright \id = f,
\end{equation}
where $f\in\mathcal A$. Then consistency requirements impose
\begin{equation}
\Delta p_\mu = \mathcal F (\Delta_0 p_\mu) \mathcal F^{-1}=\mathcal D_\mu(p\otimes\id, \id\otimes p),
\end{equation}
in accordance to Eq.~(\ref{MomentumCoproduct}).

The undeformed phase space generated by $x_\mu$, $p_\mu$ has the structure of a Hopf algebroid~\cite{IJMPA1305, Lu}, while the appropriate structure of the deformed phase space generated by $\hat x_\mu$, $p_\mu$ is that of a twisted Hopf algebroid, defined by the twist element $\mathcal F$ in equation \eqref{twist1}~\cite{IJMPA1305, SIGMA2014, PLA2013} This is a generalization of the structures of Hopf algebras.~\cite{LukierskiSkoda}.

\section{Lorentz symmetry}

The deformed phase space structures introduced above are not necessarily Lorentz-invariant in a naive way. The introduction of the energy scale $M$ into the structure of space time/phase space is incompatible with standard Lorentz symmetry, but, as is well-known after a few decades of studies of noncommutative spacetimes, this does not imply that the relativistic equivalence between inertial frames is lost. In many cases, a `deformed'  action of the Lorentz group allows to restore relativistic invariance~\cite{amelino2012fate,GoldenRulesPaper,GoldenRulesPaper2} (another type of noncommutativity involving spin, not considered here, allows to leave Lorentz symmetry undeformed~\cite{SpinNoncommutativity1,SpinNoncommutativity2,SpinNoncommutativity3}. In the present paper we are only concerned with the orbital part of the Lorentz group). A common assumption in this framework is that the Lorentz group itself (\emph{i.e.} the commutation relations between the $M_{\mu\nu}$ generators) is not deformed (this is justified by the fact that a dimensionful parameter like $M^{-1}$ cannot enter the algebraic structure of the Lorentz group). What is deformed is the action of the group on noncommutative coordinates $\hat x_\mu$ and momenta $p_\mu$, and on composed momenta (see below). Typically, there are several possible realizations of the action of the Lorentz group on our deformed phase space~\cite{AnnaRina}, and we want to `parametrize our ignorance' by writing the most generic one and then constraining its free parameters. To this end, we employ the trick of introducing a nonlinear realization of Heisenberg's algebra by performing a momentum-dependent similarity transformation on $\mathcal H$:\footnote{This trick has been recently employed in the perturbative analysis of~\cite{GoldenRulesPaper2}, which concentrated on the possible deformed kinematical structures, independently of the underlying noncommutative space.} \begin{equation}\label{DefinitionOfPiAndX}
P_\mu = \Sigma_\mu(p), ~~~ X_\mu = x_\alpha  \Psi^\alpha{}_\mu(p) + h_\mu(p),
\end{equation}
if the functions $\Sigma_\mu(p) $ and $\Psi^\alpha{}_\mu(p)$ are constrained by
\begin{equation}
\begin{aligned}
&\frac{\partial \Sigma_\mu}{\partial p_\alpha} \Psi^\alpha{}_\nu = \eta_{\mu\nu},
\\
 &\Psi_{\gamma\mu}  \frac{\partial \Psi^\beta{}_\nu}{\partial p_\gamma}  -  \Psi_{\gamma\nu}  \frac{\partial \Psi^\beta{}_\mu}{\partial p_\gamma}  = 0,
\end{aligned}
\end{equation}
in order that the basis $(P_\mu, X_\nu)$ generates an undeformed Heisenberg algebra, $[P_\mu ,P_\nu] = [ X_\mu , X_\nu] =0$, $[P_\mu , X_\nu] = - i \eta_{\mu\nu}$.

Because the new basis satisfies undeformed commutation relations, we can introduce a generator of infinitesimal Lorentz transformations as
\begin{equation}
M_{\mu\nu} = X_\mu P_\nu  - X_\nu P_\mu,
\end{equation}
and we will be ensured that the commutation relations between $M_{\mu\nu}$ and itself will close an $so(3,1)$ algebra, and their action  on $X_\mu$ and $P_\mu$ will be undeformed. However,  the action of $M_{\mu\nu}$ on $\hat x_\mu$ and $p_\mu$ is given by:
\begin{equation}\label{LorentzTransformationRule}
\begin{aligned}
&[ M_{\mu\nu} , p_\rho ] = p_\alpha \Phi^\alpha{}_{\mu\nu\rho} (p),
\\
&[ M_{\mu\nu} , \hat x_\rho ] = i \left( \hat x_\alpha \Gamma^\alpha{}_{\mu\nu\rho} (p) + \Xi_{\mu\nu\rho} (p)   \right),
\end{aligned}
\end{equation}
where $\Phi^\alpha{}_{\mu\nu\rho} (p)$,  $ \Gamma^\alpha{}_{\mu\nu\rho} (p)$ and $\Xi_{\mu\nu\rho} $ can be expressed in terms of $\Sigma_\mu (p)$,  $\Psi^\alpha{}_\mu$, $h_\mu(p)$, $\varphi^\alpha{}_\mu(p)$ and $\chi_\mu(p)$.

The Casimir operator is of course $\mathcal C = P_\mu P^\mu = \Sigma^2(p)$. The coproduct of the Lorentz generator is
\begin{equation}
\Delta M_{\mu\nu} = \mathcal F \left( \Delta_0  M_{\mu\nu}  \right) \mathcal F^{-1},
\end{equation}
where $\Delta_0 M_{\mu\nu} = M_{\mu\nu} ( \Delta_0 x, \Delta_0 p)$, which is not necessarily a primitive coproduct (because $M_{\mu\nu}$ depends in a complicated way on $x_\mu$ and $p_\mu$).

If we use the inverse relations of \eqref{DefinitionOfPiAndX} and express the generators $\hat x_\mu$ in terms of $X_\mu$, $\Sigma_\nu$, we call $\hat x_\mu (X,P)$ the `natural realization'.

\section{Constraints on the Lorentz sector}

$\kappa$-Poincar\'e is a Hopf algebra first introduced in~\cite{lukierski1991q}, and it is the most-studied Hopf-algebra deformation of relativistic symmetries. This algebra fits within our general framework (see our second example below).
$\kappa$-Poincar\'e possesses a so-called `bicrossproduct structure'~\cite{MajidRuegg}, meaning that both the algebra and the coalgebra are a semidirect product of a momentum sector with the Lorentz algebra, 
and this structure implies the existence of a co-action (or `backreaction') of the momentum sector on the Lorentz part, which is a novelty of the model. The discovery of the bicrossproduct structure was instrumental in identifying the noncommutative spacetime this Hopf algebra acts covariantly on, the so-called $\kappa$-Minkowski spacetime~\cite{MajidRuegg}. Later the phenomenon of `backreaction' was given a physical interpretation~\cite{GubitosiMercati}: it is the fact that, to boost in a covariant way a set of particles participating in a vertex, one needs to transform each particle momentum with a different rapidity. Each rapidity will depend on the momenta involved in the vertex, with some rules defined by the co-action of $\kappa$-Poincar\'e. It was later realized~\cite{GoldenRulesPaper} that a similar momentum-dependence of Lorentz transformations is a general feature of deformed relativistic kinematics, and is not limited to the Hopf-algebraic framework of~\cite{lukierski1991q,MajidRuegg,GubitosiMercati}. In fact there is a physical reason why  such an effect is a necessity, which has to do with the fact that in our deformed kinematics both the composition law of momenta, $ \oplus$, and the Lorentz transformation rule~(\ref{LorentzTransformationRule}) are nonlinear in the momenta. Consider an infinitesimal Lorentz transformation (or rotation) of a particle momentum with rapidity parameters $\omega_{\mu\nu}$ (these are just infinitesimal antisymmetric matrices). We can calculate its explicit form, at first order in $\omega_{\mu\nu}$, by using the fact that the generators $P_\mu$ transform classically:
\begin{equation}
P'_\mu = P_\mu  + {\frac 1 2} \omega^{\rho\sigma} [M_{\rho\sigma},P_\mu] = P_\mu + \omega_\mu{}^\sigma P_{\sigma} ,
\end{equation}
then we know what the action on $p_\mu$ is:
\begin{equation}
p'_\alpha = \Sigma^{-1}_\alpha \left[
\Sigma_\mu (p) + \omega_\mu{}^\sigma \Sigma_\sigma (p)\right],
\end{equation}
which is in general a nonlinear function of $p_\mu$. Expanding this function at first order in $\omega_{\mu\nu} $:
\begin{equation}
p'_\alpha = p_\alpha + \omega^{\rho \beta} \Sigma _\beta (p)  \partial_\rho \Sigma^{-1}_\alpha (p)  =  \Lambda_\alpha (\omega_{\mu\nu} , p).
\end{equation}
The relation above can be found explicitly, at first order in $M^{-1}$ and all orders in the rapidity parameter, see~\cite{DeformedLorentzTransformPaper}. By introducing the function $\Lambda$ we are seeing the infinitesimal Lorentz transformation as a map from the rapidity parameters $\omega$ and the momenta $p$ to the momenta $p'$. 
In general, such a transformation rule will not leave the generalized addition law of momenta invariant:
\begin{equation}\label{Backreaction1}
\Lambda( \omega, k \oplus q ) \neq  \Lambda( \omega, k ) \oplus \Lambda( \omega ,  q ).
\end{equation}
As observed first in~\cite{GubitosiMercati}, this is physically inconsistent,\footnote{\emph{e.g.}, it would imply that processes that are kinematically forbidden in one reference frame can become allowed in another frame, which amounts to a severe breakdown of the principle of relativity.} but can be fixed by observing that Eq.~(\ref{Backreaction1}) is implicitly assuming that the three momenta $k$, $q$ and $k \oplus q$ all transform with the same rapidity $\omega$. 
We can cure this pathology, by assuming that the rapidity, with which each momentum in Eq.~(\ref{Backreaction1}) (\emph{i.e.} $k$, $q$ and $k \oplus q$) transforms, depends on the momenta involved in the vertex:
{
\thinmuskip=0mu
\begin{equation}\label{Backreaction2}
\Lambda( \omega, k \oplus q) =  \Lambda( \omega^{(1)}(k,q), k) \oplus \Lambda( \omega^{(2)}(k,q), q).
\end{equation}}
The equations above impose a series of constraints on the functions $\omega^{(1)}{}_\mu{}^\nu(k,q)$, $\omega^{(2)}{}_\mu{}^\nu(k,q)$, $\mathcal D_\mu (k,q)$ and $\Lambda_\mu( \omega,q)$.
In~\cite{GoldenRulesPaper} these constraints were calculated in a perturbative setting (at first order in $M^{-1}$ and assuming undeformed rotational symmetry). The constraints found were enough to completely fix $\omega^{(1)}{}_\mu{}^\nu(k,q)$ and $\omega^{(2)}{}_\mu{}^\nu(k,q)$, and to establish a few relationships between the parameters of $\mathcal D_\mu (k,q)$ and $\Lambda_\mu(\omega,q)$.\footnote{Alternatively, these constraints could be interpreted as fixing completely the parameters of $\Lambda_\mu(\omega,q)$ as functions of the parameters of $\mathcal D_\mu (k,q)$, $\omega^{(1)}{}_\mu{}^\nu(k,q)$ and $\omega^{(2)}{}_\mu{}^\nu(k,q)$.}

\section{Examples}

For Snyder-type spaces~\cite{Battisti, Mignemi}
\begin{equation}
\hat x_\mu = x_\mu \varphi_1(p^2) + (x\cdot p) p_\mu \varphi_2(p^2) + p_\mu \chi(p^2),
\end{equation}
the relativistic addition of momenta is covariant under standard Lorentz transformations with no momentum dependence of Lorentz transformations: $\omega^{(1)}=\omega^{(2)}=\omega$.

For general $u_\mu$ vector-like deformations of Minkowski space at first order in $M^{-1}$, the realization is given by:
\begin{equation}\begin{split}
&\hat x_\mu = x_\mu + c_1 x_\mu (u\cdot p) + c_2 u_\mu(x\cdot p) +
\\
&c_3 u_\mu (u\cdot x)(u \cdot p) + c_4 (u\cdot x)p_\mu +\okk.
\end{split}\end{equation}
where $u^2\in\{-1,0,1\}$.

The commutator of coordinates is of the form
\begin{equation}
[\hat x_\mu, \hat x_\nu] = i(a_\mu \hat x_\nu - a_\nu \hat x_\mu) + \okk
\end{equation}
where $a_\mu = (c_1-c_2)u_\mu$.

The basis $X_\mu$, $P_\mu$ is given by:
\begin{align}
\begin{split}
X_\mu =  x_\mu &- d_1 x_\mu (u\cdot p) - d_1 u_\mu(x\cdot p) \\
& -2d_3 u_\mu (u\cdot x)(u \cdot p) - 2d_4 (u\cdot x)p_\mu \\&+ \okk
\end{split} \\
\begin{split}
P_\mu = p_\mu &+ d_1(u\cdot p)p_\mu + d_2 u_\mu p^2\\
&+ d_3 (u\cdot p)^2 u_\mu +\okk
\end{split}
\end{align}

In 1+1 dimensions, the rapidities $\omega^{(1)}(k,q)$ and $\omega^{(2)}(k,q)$ are
\begin{equation}\label{BackreactionVectorLike}
\begin{split}
\omega^{(1)}(k,q)&= \left[ 1-(c_2+d_1)u\cdot q \right] \omega, \\
\omega^{(2)}(k,q)&= \left[ 1-(c_1+d_1)u\cdot k \right] \omega.
\end{split}\end{equation}
(there is only one possible infinitesimal Lorentz transformation in $1+1$ dimensions, it is the boost in the $1$-direction, and therefore the rapidity parameter has no indices). To further clarify the physical meaning of these expressions, Eq.~(\ref{BackreactionVectorLike}) gives the rapidity $\omega^{(1)}(k,q)$ with which the momentum $k$ transform, and the rapidity $\omega^{(2)}(k,q)$ with which momentum $q$ transform, given the rapidity $\omega$ with which the composed momentum $k \oplus q$ transform, if $k$, $q$ and $k\oplus q$ participate in a trivalent vertex, according to formula~(\ref{Backreaction2}).

In the $\kappa$-Poincar\'e case, the $d$'s are given in terms of the $c$'s by
\begin{equation}
d_1=-c_2, ~ d_2=-\frac{c_1-c_2+c_4}2, ~ d_3=-\frac{c_3}2.
\end{equation}
The rapidities $\omega^{(1)}(k,q)$ and $\omega^{(2)}(k,q)$ then simplify to (again in $1+1$ dimensions)
\begin{equation}\begin{split}
\omega^{(1)}(k,q)&=\omega, \\
\omega^{(2)}(k,q)&=\left( 1-a\cdot k \right) \omega,
\end{split}\end{equation}
which coincides with the results found in~\cite{GubitosiMercati} and related references.

\section{Outlook and discussion}

By  using general nonlinear redefinitions of the basis of the Heisenberg algebra (with the only assumption of being up-to-linear order in the coordinates $x_\mu$), we were able to encompass within a unified framework a large class of noncommutative spacetimes. Within this framework, we have a prescription to uniquely determine the effect of combining two plane waves, which gives a composition rule for momenta that deforms the momentum conservation laws of special relativity into  a nonlinear operation. Moreover, in our framework, we have unique prescriptions that characterize the momenta and determine the Hopf algebroid structure. These algebraic structures have been recognized to be the suitable framework to describe the symmetries of noncommutative spacetimes. Finally, we studied how to introduce Lorentz symmetry in our framework: this cannot be done uniquely, as it introduces a certain degree of arbitrariness (in a perturbative approach this amounts to a few free parameters). However, this arbitrariness is constrained by the requirement of relativistic invariance of the momentum composition law. The recently-discovered phenomenon of  momentum-dependence of Lorentz transformations~\cite{GubitosiMercati}  (\emph{i.e.} the rapidity with which momenta participating to a vertex transform depends on the momenta) was found to constrain the form of the acceptable Lorentz transformations. To show the applicability of the framework we presented, we concluded this Letter with two examples: one is a class of noncommutative spacetimes which generalizes that introduced by Snyder in 1947~\cite{Snyder}. This example leads to no momentum-dependence effect and no deformation of the Lorentz group action. The second one is a generalization of the so-called $\kappa$-Minkowski spacetime, which is the first one for which the momentum-dependence effect was discovered. The phenomenon of momentum-dependence of Lorentz transformations could have interesting  phenomenological consequences. For example, in astrophysical settings one can have high-energy particles emitted from high-redshift sources. This is a situation in which both the momenta involved in a process and the rapidity identifying the reference frame are appreciably large (in Planck units), and the momentum-dependence effect might become manifest.

\section*{Acknowledgements}
The work by S.M. and D.P. has been fully supported by Croatian Science Foundation under the Project No. IP-2014-09-9582 as well as by the H2020 Twinning project No. 692194, ``RBI-T-WINNING''. F.M. was supported by Perimeter Institute for Theoretical Physics and by a Marie Curie fellowship of the Istituto Nazionale di Alta Matematica. Research at Perimeter Institute is supported by the Government of Canada through Industry Canada and by the Province of Ontario through the Ministry of Research and Innovation.


\bibliographystyle{elsarticle-num.bst}
\bibliography{bibNCspaces}

\end{document}